\documentclass[]{tPHM2e}

\def \etal {{\it et al.\ }}

\def \ur {URu$_2$Si$_2$\ }
\newcommand{\comment}[1]{}

\begin{document}
\doi{10.1080/14786435.20xx.xxxxxx}
\issn{1478-6443}
\issnp{1478-6435}
\jvol{00} \jnum{00} \jyear{2010} 

\markboth{Taylor \& Francis and I.T. Consultant}{Philosophical Magazine}

\title{{\itshape Optical study of hybridization and hidden order in \ur}}

\author{J.S. Hall$^{\rm a}$ and T. Timusk$^{\rm a,b}$$^{\ast}$\thanks{$^\ast$Corresponding author. Email:timusk@mcmaster.ca
\vspace{6pt}}\\\vspace{6pt}  $^{\rm a}${\em{Department of Physics and Astronomy, McMaster University, Hamilton ON Canada}}; $^{\rm b}${\em{Canadian Institute for Advanced Resarch}}\\\vspace{6pt}\received{Received xxx.xx.xx} }

\maketitle

\begin{abstract}
We summarize existing optical data of \ur to clarify the nature of the hidden order transition in this heavy fermion metal.  Hybridization develops between 50 K and 17.5 K, and a coherent Drude peak emerges which mirrors the changes in the dc resistivity.  The Drude weight indicates that there is little change in the effective mass of these carriers in this temperature range.  In addition, there is a flat background conductivity that develops a partial hybridization gap at 10 meV as the temperature is lowered, shifting spectral weight to higher frequencies above 300 meV.  Below 30 K the carriers become increasingly coherent and Fermi-liquid-like as the hidden order transition is approached. The hidden order state in \ur is characterized by multiple anisotropic gaps.  The gap parameter $\Delta_a=3.2$ meV in the {\it ab}-plane. In the {\it c}-direction, there are two distinct gaps with magnitudes of $\Delta_{c1}=2.7$ meV and $\Delta_{c2}=1.8$ meV. These observations are in good agreement with other spectroscopic measurements.  Overall, the spectrum can be fit by a Dynes-type density of states model to extract values of the hidden order gap. The transfer of spectral weight strongly resembles what one sees in density wave transitions.

\end{abstract}

\section{Introduction}

Optical spectroscopy is an ideal tool for studying the low-energy electrodynamics of strongly correlated metals such as URu$_2$Si$_2$. It is a bulk probe, penetrating hundreds of atomic layers into the material; it is therefore not sensitive to surface states. Furthermore, it allows the study of cut and polished surfaces rather than being restricted only to the cleavage plane in the manner of STM/STS and ARPES. Optical measurements can access a very wide range of energies, from the far-infrared ($\sim$ 2 meV) to the ultraviolet ($\sim$ 5 eV). They yield direct high resolution ($ < $ 0.10 meV) spectroscopic information about the energy gaps of ordered states and can be used to extract the spectrum of excitations responsible for the self energy of the free carriers. Optical sum rules can be used to distinguish states by determining where the spectral weight is lost in the gapped conductivity and to where it is transferred: in the zero frequency condensate state for superconductors, just above $2\Delta$ for density waves, or to high frequencies for strong correlation gaps. In \ur  the charge carrier dynamics appears to be strongly affected by the hidden order state, and optical spectroscopy is an important tool for understanding how the charge carrier dynamics evolves from the incoherent state into the ordered state.

Shortly after the first characterization of \ur by transport measurements \cite{palstra85, palstra86,maple86}, reflectance measurements were performed \cite{bonn88} that revealed the presence of a strong absorption band centred around 5 meV in the ordered phase below  $T_{HO} =$ 17.5 K.  Kramers-Kronig analysis of the reflectance showed that the absorption band was the result of the gradual opening of a gap in the conductivity below the hidden order temperature. The spectral weight lost in the gap was transferred to frequencies just above the gap, typical of a density wave transition.  Subsequent optical work has centred around investigations of the effects of doping on the hidden order \cite{stthieme95,degiorgi97}, of the effect of hybridization on the electrodynamics \cite{levallois11,liu11,nagel12,guo12} and the anisotropy of the hidden order parameter \cite{hall12}. The superconducting state occurs below the temperatures available in typical optical experiments; likewise, the large moment antiferromagnet state (LM-AFM) induced by hydrostatic pressure has not yet been investigated optically.\\

\begin{figure*}
\begin{center}
\vspace*{-0.5 cm}
\hspace*{-0.8cm}
\resizebox*{10cm}{!}{\includegraphics{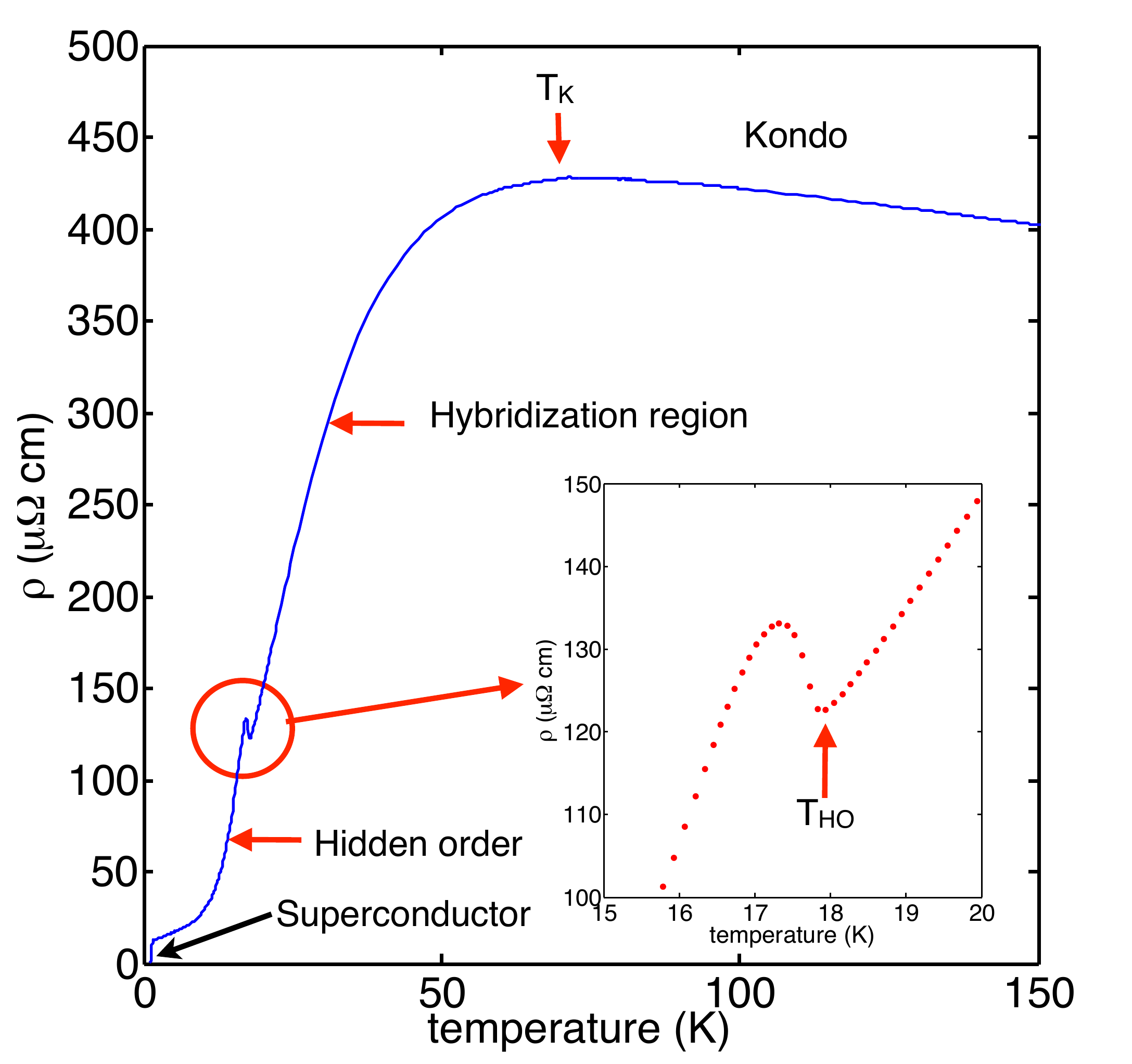}}
\vspace*{0.0 cm}%
\caption{(color online): The resistivity of \ur from \cite{butch10}.  Above $T_K = 70$ K the resistivity is Kondo-like in that it increases as the temperature is lowered.  Below $T_K$ the resistivity drops dramatically as the uranium $f$ electrons begin to hybridize with the $s$ and $p$ electrons.  The hidden order transition takes place at $T_{HO}= 17.5 $ K  and the resistivity rises slightly initially but then drops dramatically as the hidden order gap opens. Finally, at $T_C$ $\sim 1.5$ K, \ur becomes a superconductor. The inset shows the resistivity near the hidden order transition.   }
\label{fig:dc.res}
\vspace*{0.0cm}
\end{center}
\end{figure*}

Figure 1 shows the dc resistivity of \ur  \cite{palstra86} as a function of temperature. Four separate regions can be identified.  At high temperatures (above 100 K) is the Kondo region where the resistivity rises as the temperature is lowered, a behaviour typical of other heavy fermion systems in which the uranium $f$ electrons are thought to act as localized magnetic impurities that cause large incoherent scattering. At T$_{K}$ $\sim$ 70 K the resistivity begins to drop as the scattering rate begins to decrease rapidly.   At the hidden order transition $T_{HO} = 17.5$ K there is a small but distinct  jump in the resistivity. In the hidden order state the resistivity falls further with decreasing temperature, finally falling to zero as the superconducting state is reached, at 1.5 K in pure samples.

In the following, we focus on two regions shown in figure \ref{fig:dc.res}: first, between $T_K$ and $T_{HO}$ where the coherent conductivity first appears,  labelled the hybridization region; and second, the region where the hidden order gap appears. We will not discuss the region above $T_K$  where the optical conductivity, shown in figure \ref{fig:sigma.fl}, resembles many other heavy electron systems \cite{marabelli86,sulewski86,dordevic01, mena03, singley02, sichelschmidt06} in their incoherent high temperature state.  Here the conductivity is frequency and temperature independent and has a magnitude that approaches the Mott-Ioffe-Regel limit wherein the mean free path of the carriers approaches the interatomic spacing.  Also, we will not address the superconducting state, which is not readily accessible to infrared reflectance measurements.

\section{Hybridization region}

\begin{figure*}
\begin{center}
\vspace*{-0.5 cm}
\hspace*{-0.8cm}
\resizebox*{10cm}{!}{\includegraphics{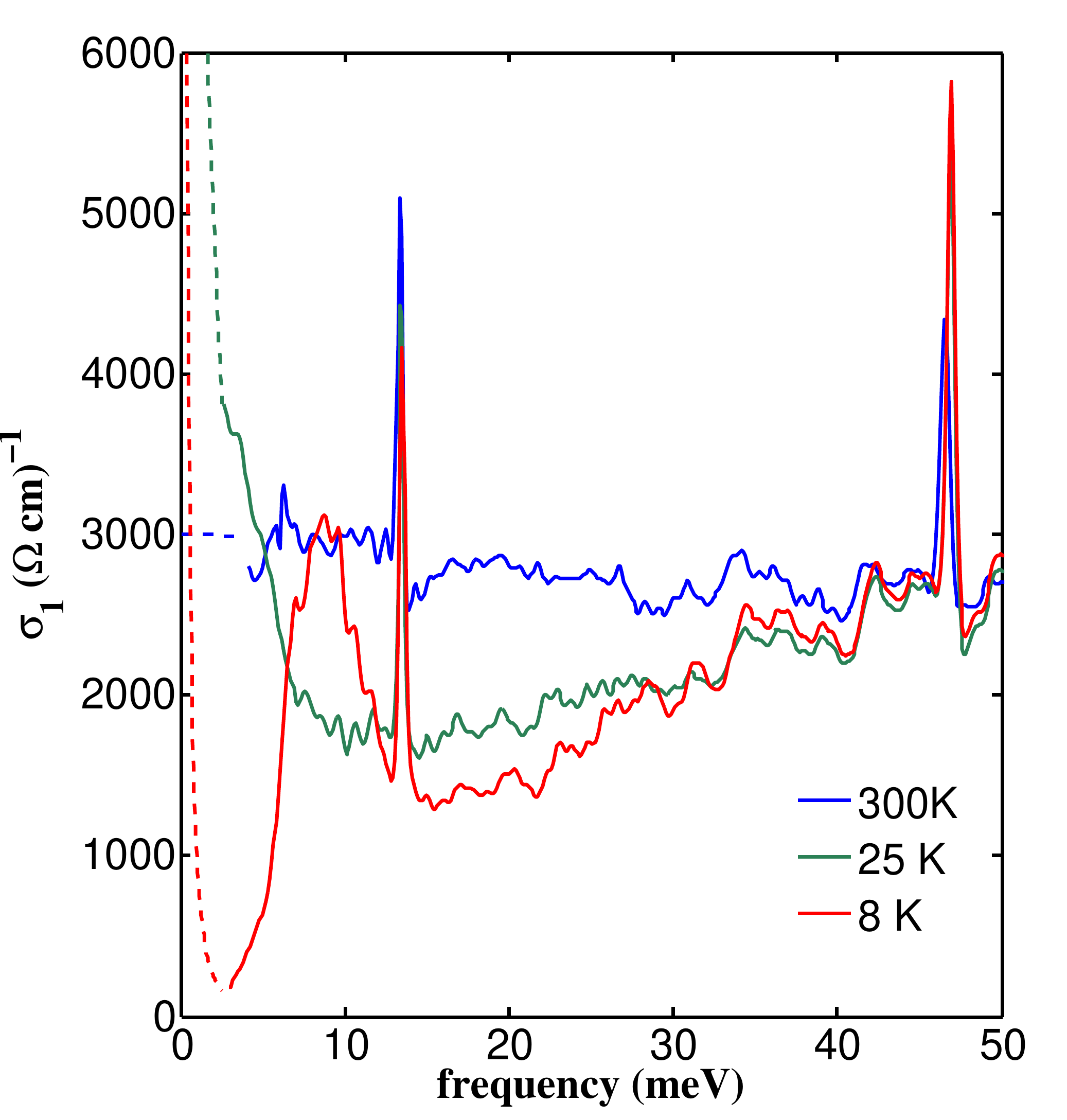}}
\vspace*{0.0 cm}%
\caption{(color online): The optical conductivity as a function of photon energy in the region dominated by the hybridization gap. Below 4 meV the conductivity has been fitted to a Drude peak whose amplitude agrees with the dc resistivity. Below 10 meV the Drude weight dominates. The spectral weight in the hybridization gap region, 5 to 40 meV, is lost to higher frequencies. The sharp peaks at 13.5  and 46.9 meV are optically active phonons. }
\label{fig:sigma.fl}
\vspace*{0.0cm}
\end{center}
\end{figure*}

Optical studies of the temperature regime between approximately $T_{K}$ and $T_{HO}$ reveal a great deal about the evolution of the electrodynamics as the heavy fermion state forms. This is shown in figure \ref{fig:sigma.fl}, where the optical conductivity is plotted as a function of frequency at three temperatures.  As the temperature is lowered the single flat background that dominates at 300 K develops a Drude peak at low frequency and a broad gap-like depression centred at 10 meV.  The c-axis conductivity is rather higher than that of the a-axis, in keeping with transport measurements \cite{hall12}, but is similar in overall features to what is shown in figure \ref{fig:sigma.fl}.  In the heavy Fermion literature this gap-like depression has been identified as the hybridization gap that is expected to develop as the localized $f$ electrons hybridize with the mobile $s$ and $p$ electrons \cite{millis87}. However, both STM \cite{schmidt10,aynajian10} and ARPES \cite{santandersyro09, yoshida10,boariu13,chatterjee13}  measurements suggest that just above the HO transition there are light carriers and that hybridization may not be complete. As figure \ref{fig:sigma.fl} shows, the conductivity at 10 meV has dropped to half its high temperature limiting value and a gap is not completely formed at 25 K.  Similar results have been obtained by other investigators \cite{levallois11,guo12}.

With decreasing temperature the Drude peak narrows and increases in height as the scattering becomes increasingly coherent and the dc resistivity decreases.  If the {\it only} change were the development of heavy mass we would expect to see the Drude peak narrow with the lost spectral weight shifted into a Holstein sideband \cite{grimvall81}, but the height of the Drude peak would remain the same. Clearly, this is not the case in URu$_2$Si$_2$. There is another process at work; the carrier mobility is increasing, which by itself would raise the height of the Drude peak but not affect its spectral weight, {\it i.e.} the area under the peak. The combination of the two effects gives a Drude peak that is narrowing and increasing in height but also losing some spectral weight to hybridization.  

This behaviour, namely the suppression of the background conductivity at higher frequencies and the sharpening of the Drude peak, has been observed in other heavy fermion compounds, such as YbFe$_4$Sb$_{12}$ and CeRu$_4$Sb$_{12}$ \cite{dordevic01}, CeCoIn$_{5}$ \cite{singley02}, MnSi \cite{mena03}, CaFe$_4$Sb$_{12}$ and BaFe$_4$Sb$_{12}$ \cite{sichelschmidt06}, and UCu$_5$ and UPd$_2$Al$_3$ \cite{degiorgi97}.

\begin{figure*}
\begin{center}
\vspace*{-0.5 cm}
\hspace*{-0.8cm}
\resizebox*{10cm}{!}{\includegraphics{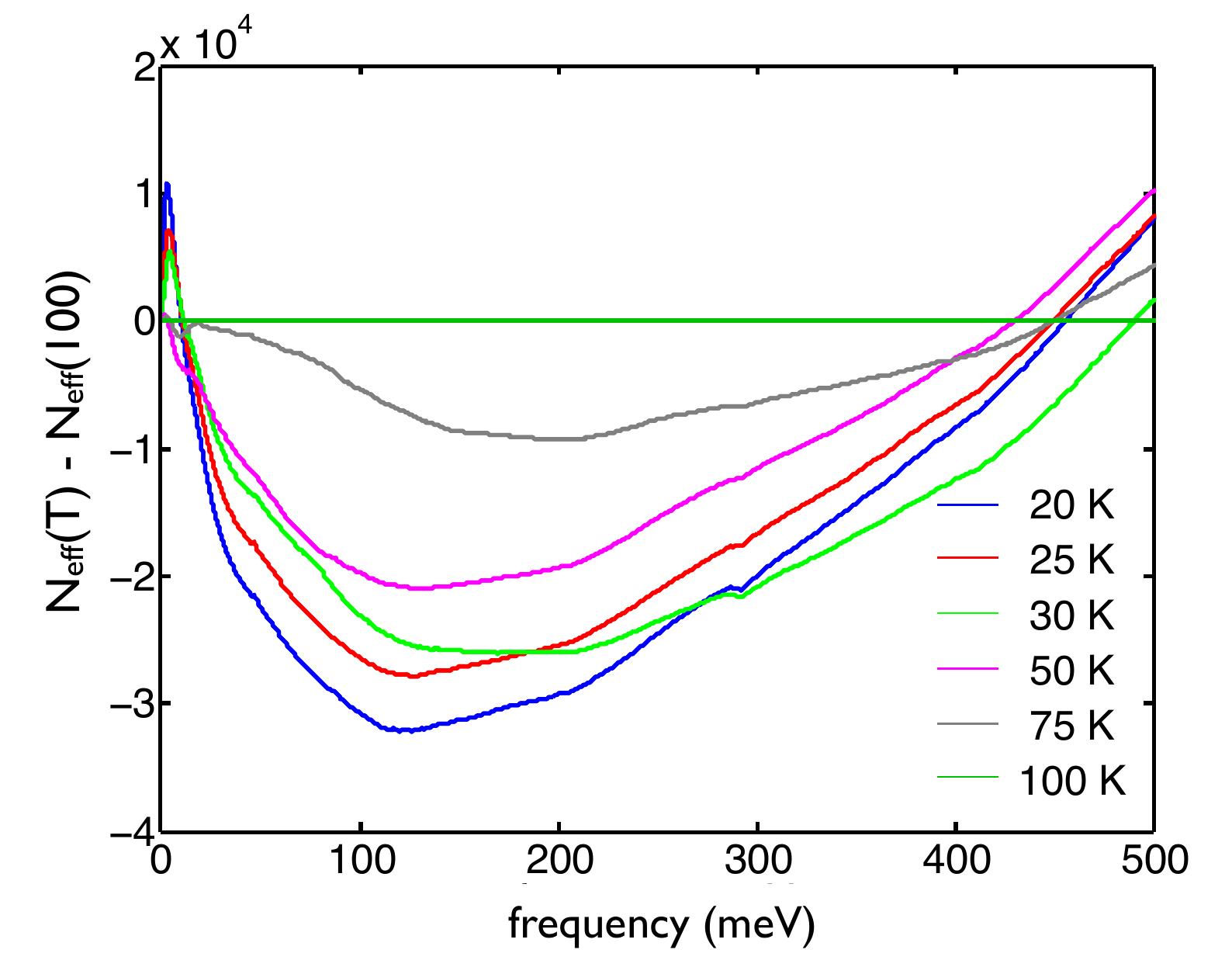}}
\vspace*{0.0 cm}%
\caption{(color online): Partial spectral weight.  The spectral weight below 10 meV  is dominated by the Drude peak, which vanishes above 30 K. The curves merge in the 10 meV region showing that the Drude weight does not change much in this temperature region.  The high frequency range is dominated by the spectral weight in the incoherent region and the curves join at 450 meV showing that the pseudogap spectral weight lost in the 10 to 100 meV region is recovered in the  100 to 450 meV region. Finally there is new spectral weight being added above 450 meV. We attribute this to interband absorption.}
\label{fig:sw}
\vspace*{0.0cm}
\end{center}
\end{figure*}

Figure \ref{fig:sw} shows the partial spectral weight, $N_{eff}(\Omega,T)-N_{eff}(\Omega,100)$,  where $N_{eff}(\Omega,T)=\int_{0}^{\Omega}\sigma(\omega,T)d\omega$ and where we  have subtracted the partial spectral weight at 100 K over a wider range of frequencies. We see a large loss of spectral weight between 10 meV and 100 mev, and a recovery in the region from 100 to 450 meV corresponding to the formation of  the hybridization gap.  Very similar data has been presented by Guo \etal\cite{guo12}. In the context of conventional electron-boson interaction models \cite{grimvall81} this energy scale would imply coupling to excitations with energies of the order of 300 meV, much higher than the Kondo scale $T_{K} \approx 6$ meV. One should note, however, that measurements of spectral weight recovery in the high frequency region are difficult in view of the small changes in the optical conductivity and the possibility of overlapping interband absorption in this region.  For example, instead of the expected flattening of the curves above 400 meV, we see a crossing to positive values.  In agreement with this view the optical reflectance shows a clear interband feature at 400 meV \cite{hall12}.

It is difficult to estimate the spectral weight of the Drude peak accurately, but in figure \ref{fig:sw} the narrow maximum near zero frequency  is the Drude contribution, about a fifth of the large negative arising from the hybridization gap. It was suggested in Nagel \etal that the total spectral weight in the Drude peak is conserved \cite{nagel12}, as suggested by the crossover of the partial spectral weight curves at 10 meV. One cannot rule out {\it some} changes in Drude weight between $T_K$ and $T_{HO}$, but any change within the limits of the data would be far too small to explain the large mass seen in the specific heat just above the hidden order transition where $m^*/m=25$\cite{maple86}. Fitting a Drude peak to the high frequency tail of the conductivity down to 2.0 meV (the limit of our data) and the measured dc conductivity, we find little evidence for mass changes above a factor of two.  Similar measurements by Levallois \etal\cite{levallois11} show only a slightly larger change in the Drude weight between 90 and 20 K.  Another way of estimating the mass change in this region is to use the expression $m^*/m = (V_K/k_BT_K)^2$ where $V_K$ is the zero temperature value of the hybridization gap and $T_K=70$ K, the temperature where the hybridization starts \cite{millis87,dordevic01}. Since $V_K$ is ill defined in our data, we will use the $m^*/m=2.5$ from ref. \cite{levallois11} to find $V_K=10$ meV.  This value for the hybridization gap agrees roughly with the flat region of minimum conductivity in Fig. 2 between 5 and 15 meV.

Figure \ref{fig:sw} shows only the portion of the spectral weight that is changing between 100 K and the temperature $T$.  There remains a background incoherent conductivity that does not change in this range of temperature.  It is difficult to estimate the spectral weight of this component but if it extends up to 0.6 eV it could well explain the $\gamma$ value of the specific heat. As we will see below, it is this incoherent component that hybridizes at $T_{HO}$.  The original spectroscopy paper of Bonn \etal cited an effective mass of 40 $\sim$ m$_{e}$ obtained from the limited spectral region available by fitting a Lorentz oscillator to the high frequency data and using this to estimate the plasma frequency, which as a result includes both the Drude component and the incoherent background in the total spectral weight.

\begin{figure*}
\begin{center}
\vspace*{-0.5 cm}
\hspace*{-0.0cm}
\resizebox*{14cm}{!}{\includegraphics{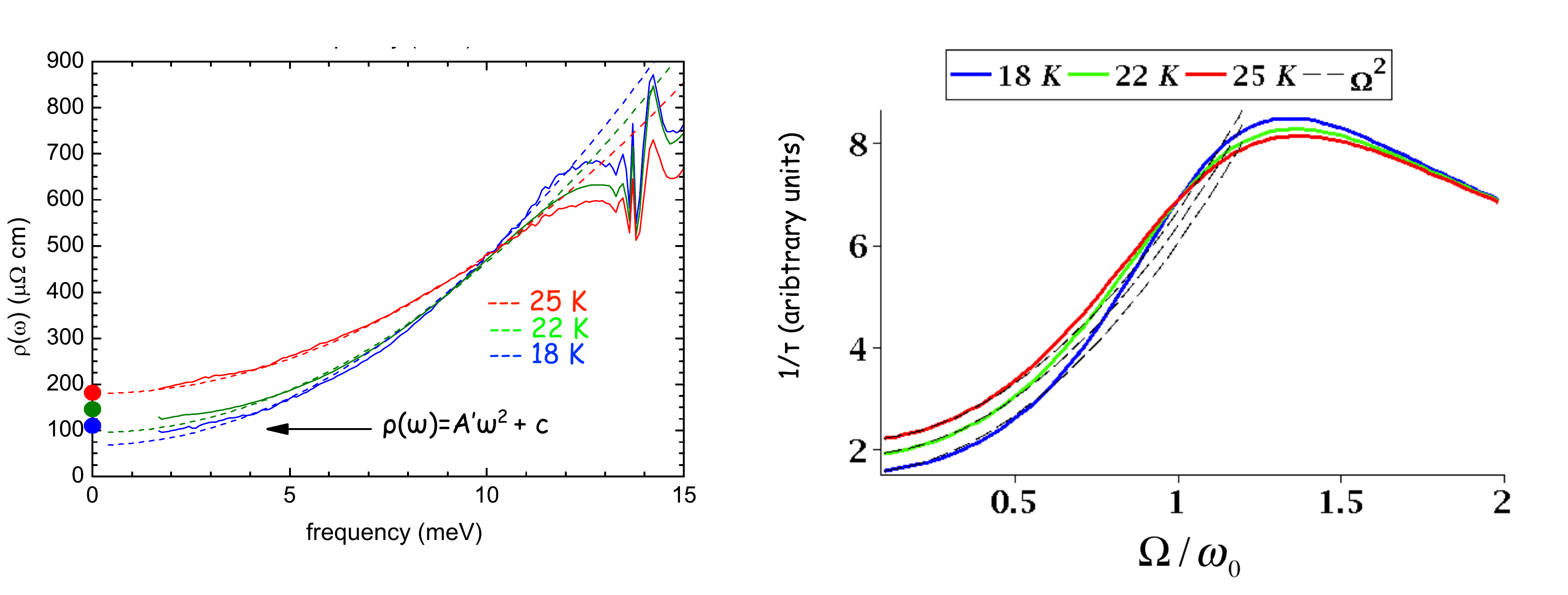}}
\vspace*{-0.8 cm}%
\caption{(color online): The left panel shows the low-energy frequency dependent resistivity $\rho(\omega)$ of \ur. The experimental curves (solid lines) are compared to a Fermi liquid fit (dashed lines) with the coefficient $A'$ and an offset $c(T)$ determined by a least squares fit to the experimental data. The right panel shows the frequency dependent scattering rate calculated for resonant impurity scattering, with dashed lines showing the fit to Fermi liquid scattering. After U. Nagel {\it et al.}, PNAS {\bf 109}, (2012( p.19161, and D.L. Maslov and A.V. Chubukov, Phys. Rev. B {\bf 86}, (20112) 155137.}
\label{fig:rho.opt}
\vspace*{0.0cm}
\end{center}
\end{figure*}

Figure \ref{fig:rho.opt} shows the optical resistivity $\rho(\omega)={\cal R}{e}(1/\sigma(\omega))$ from Nagel \etal \cite{nagel12}.  Below 22 K it varies quadratically with frequency (with slight deviations at the lowest frequencies, presumably the result of impurities in the sample raising the dc resistivity) which indicates the onset of purely Fermi-liquid-like behaviour. The resistivity of a Fermi liquid, where electron-electron umklapp scattering is the dominant mechanism, is the sum of two terms, one quadratic in frequency and the other quadratic in temperature\cite{gurzhi59}:
\begin{eqnarray}
\rho(\omega,T) = A'(\hbar\omega^2 + b\pi^2 (k_BT^2))
\end{eqnarray}
where the coefficient $b=4$. This is not what is observed in URu$_2$Si$_2$. The observed ratio of the temperature and frequency coefficient $b$ is 1.0$\pm$0.1 \cite{nagel12}. Similar deviations from simple electron-electron scattering formula have been reported by Sulewski \etal\cite{sulewski86} in UPt$_3$.  A search of the literature shows that virtually all correlated electron materials have a $b < 4$ and that the coefficient $b$, where it has been measured, varies from material to material from less than one up to 2.5. One way this can be understood  is in terms of a suggestion of Maslov and Chubukov\cite{maslov12} of scattering from resonant impurities leading to quadratic frequency and temperature dependence rather than inelastic electron-electron scattering. The coefficient {\it b} then lies along a continuum from 1 to 4, with the value determined by the relative strength of the elastic and inelastic scattering. An obvious source of the elastic scattering centres are the un-hybridized uranium f-electrons.

\begin{figure*}
\begin{center}
\vspace*{-0.5 cm}
\hspace*{-0.8cm}
\resizebox*{10cm}{!}{\includegraphics{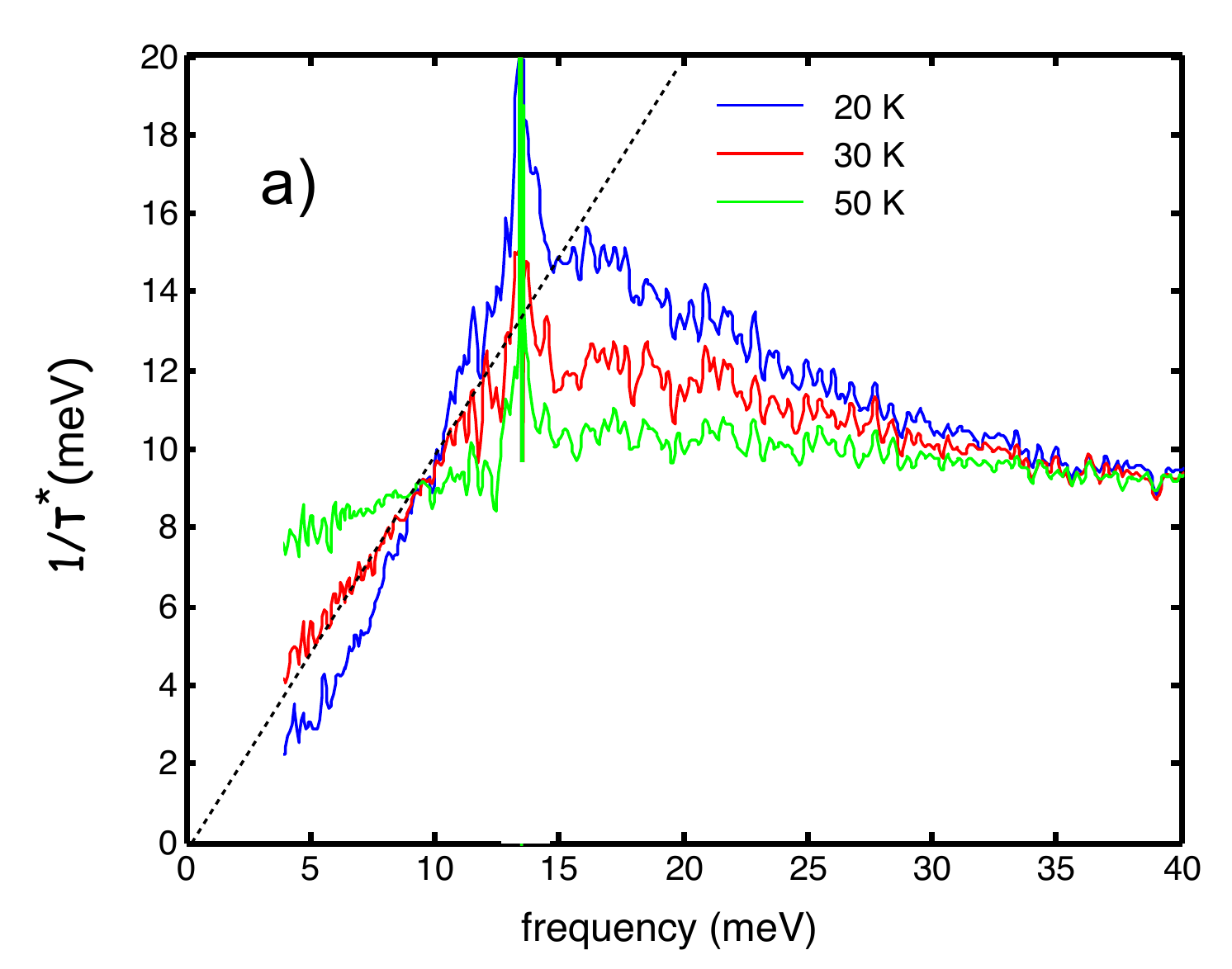}}
\vspace*{0.0 cm}%
\caption{(color online): The frequency dependent renormalized scattering rate $1/\tau^*$ at three temperatures.  As the temperature is raised the Fermi liquid scattering at 20 K is gradually replaced by an incoherent background. Coherent quasiparticles exist below the dashed line $\omega=1/\tau*$.  This condition is satisfied for $T = 30$ K. After U. Nagel {\it et al.}, PNAS {\bf 109}, (2012) p.19161.}
\label{fig:tau}
\vspace*{0.0cm}
\end{center}
\end{figure*}

The optical resistivity $\rho(\omega,T)$ can be converted to the scattering rate if the plasma frequency is known.  We plot in figure \ref{fig:tau}  the renormalized scattering rate $1/\tau^*$ based on the Drude plasma frequency of 418 meV at three temperatures from Nagel \etal\cite{nagel12}.  At 50 K the scattering rate is flat and does not vary with frequency, a signature of incoherent hopping transport.  The dashed line corresponds to the condition $\omega=1/\tau^*$ for coherent transport.  One can then draw the conclusion that coherent transport of the Drude component of the carriers starts below 30 K.  This estimate of quasiparticle lifetime can be compared with time resolved ARPES scattering from Dakovski \etal who find a quasiparticle scattering rate of 2.2  mev ($\tau=301$ fs) in the hidden order state located near the Fermi surface "hotspots".  We can identify these quasiparticles with our Drude carriers that have even longer lifetimes at low temperatures.  Just above the hidden order transition Dakovski \etal find short-lived quasiparticles with scattering rates of 15 meV (44 fs) at 19 K. Presumably these correspond to the carriers contributing to the incoherent background that hybridizes at 17.5 K.

\begin{figure*}
\begin{center}
\vspace*{-0.5 cm}
\hspace*{-0.8cm}
\resizebox*{10cm}{!}{\includegraphics{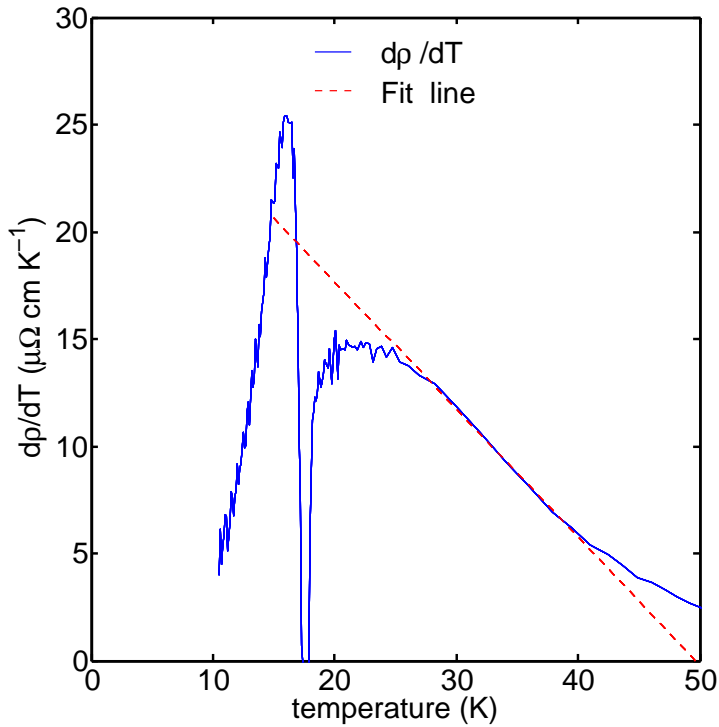}}
\vspace*{0.0 cm}%
\caption{(color online): The rate of change of dc resistivity with temperature $d\rho(T)/dT$\cite{butch10} (solid line). The dashed line shows the trend above 30 K.  There is a component that grows below 30 K. As shown in Nagel \etal\cite{nagel12}, in this region the transport approaches Fermi liquid behaviour.  Within 1 K of the transition this component grows more rapidly.  The derivative goes to zero at 17.85 K which we suggest is the hidden order transition, T$_{HO}$.}
\label{fig:drho}
\vspace*{0.0cm}
\end{center}
\end{figure*}

In addition to the onset of coherence seen in  the Drude scattering rate, there is further evidence for changes in the electronic properties at T $\approx$ 30 K, well above the hidden order temperature. These features have been attributed to a "pseudogap" in analogy with the cuprate superconductors where similar signatures are seen. NMR $(T_1T)^{-1}$ \cite{shirer13} shows a depression growing in depth below 30 K. No new optical spectral features are seen at this temperature: the hybridization gap measured in optics develops continuously from 70 K to 17.5 K, as seen by Nagel \etal \cite{nagel12}, and any precursors to the hidden order gap would be expected to appear only at very low frequency, below 2 meV.  According to Levallois \etal\cite{levallois11} there is a break in the rate of change of the Drude plasma frequency at 30 K as well as the scattering rate -- both evolve more rapidly below this temperature.  But the dc resistivity does not show any discontinuous kinks between 35 K and 17.5 K, only a gradual flattening of the curve as $T_{HO}$ is approached from above, followed by an upturn below the transition. As shown in figure \ref{fig:drho} the rate of change of dc resistivity acquires a negative component below 30 K but there is no sharp kink. A gap has two opposite effects on the dc resistivity: a reduction in the number of carriers $N$ at the Fermi surface causes an increase in resistivity, but a gap also reduces the number of states available for scattering, reducing $1/\tau$.  Just below the hidden order transition it seems that the decrease in $N$ dominates, causing an upturn in resistivity. Well below the hidden order transition the resistivity drops exponentially due to the gapping of final states available for scattering.  This is similar to what happens in the high temperature superconductors below the superconducting transition temperature \cite{kamal98}. By contrast, the opposite is true at the cuprate pseudogap where at $T^*$ the resistivity drops which is evidence that the scattering reduction dominates.  

As figure \ref{fig:rho.opt} shows there are no distinct bosonic features in the optical resistivity spectra of \ur in the frequency range 2 to 10 meV, the type of excitations seen in the 40 meV range in many cuprate superconductors, and are associated with peaks in the magnetic fluctuation spectrum\cite{carbotte11}.  Instead, the excitation spectrum is smooth and has the $\omega^2$ dependence of a Fermi liquid.

Time-resolved THz reflectance measurements \cite{liu11} also offer insight into the electrodynamics in the coherent regime. Above $T_K$ a single and relatively constant fast decay in the reflectance is observed. As the temperature is lowered below 60 K, this fast decay slows down. This can be attributed to the opening of a partial hybridization gap. At 25 K the single-decay model is no longer accurate, and a two-component (fast component and slow component) decay must be fitted to the data. This is strongly suggestive of the opening of a partial gap at 25 K. The PCS measurements of Park \etal\cite{park12} see a partial gap open below 27 K, also about 10 meV, but they do not see any sharp features at the hidden order transition, and it has been suggested \cite{rodrigo97} that this may be due to the pressure of the metal tip on the sample. ARPES measurements also see a gap open at the X-point in k-space below 25 K \cite{boariu13}. This gap has a value of approximately 10 meV and is well explained by a simple hybridization model, and furthermore, is associated with a different part of the Fermi surface than the hidden order gap. All of this suggests that this is unrelated to the hidden order state.

In summary, in the state between 50 K and 17.5 K the optical conductivity is dominated by two channels. The first is a coherent Drude channel which is responsible for the changes in the dc resistivity.  There is some evidence for an increase in the effective mass of these carriers in this temperature range but not enough to explain the large specific heat coefficient.  The second channel has a flat, frequency independent conductivity that develops a partial gap at 10 meV as the temperature is lowered. The spectral weight lost in this gap is recovered at a much higher frequency of 300 meV.  This is in broad agreement with other spectroscopic measurements. 

\section{The hidden order state}

Unlike transport measurements, the optical data show no discontinuities at the onset of the hidden order transition. Instead, an absorption feature appears gradually in the reflectance, which shifts to a higher frequency as the temperature is lowered and ends up, in the low temperature limit, at $\omega \sim$ 5 meV when measured with light polarized along the a-axis and at $\omega \sim$ 4 meV when measured with light polarized along the c-axis \cite{hall12}. The absorption becomes stronger as the temperature is decreased below $T_{HO}$. No other optical response is associated with the onset of the hidden order state. The optical conductivity shows that the HO state is accompanied by a gap in the low frequency conductivity; nonetheless, the Drude peak continues to sharpen, indicating that the greatly increased carrier mobility more than compensates for the loss of carriers. The spectral weight that is lost in the gap region is recovered in a peak above the gap energy. In the c-axis conductivity there is clear evidence for the opening of a second gap with a separate, distinct energy. The larger gap in the c-axis is also at a different energy as compared with  the gap in the a-axis conductivity. These two observations indicate that the hidden order parameter is anisotropic.

It was initially assumed that the hidden order state was antiferromagnetic, or possibly a spin density wave state \cite{palstra86, maple86, schablitz86}, and the early optical measurements of the 1980s and 1990s were interpreted in this framework. Gradually, however, it became apparent that the tiny ordered moment of 0.03 $\mu_B$ per U atom was far too small to account for the specific heat data: more entropy was being quenched at the transition, on the order of R ln(2), than could be ascribed to magnetic ordering. More recently it has been argued \cite{amitsuka07} that the small moment antiferromagnet (SM-AFM) state is extrinsic, caused by strain regions in the lattice, and is not a property of the hidden order state. In addition, when hydrostatic pressure is applied to the crystal, there is a first-order phase transition to the LM-AFM state, further suggesting that the hidden order is unrelated to antiferromagnetism. 

\begin{figure*}
\begin{center}
\vspace*{-0.5 cm}
\hspace*{-0.8cm}
\resizebox*{10cm}{!}{\includegraphics{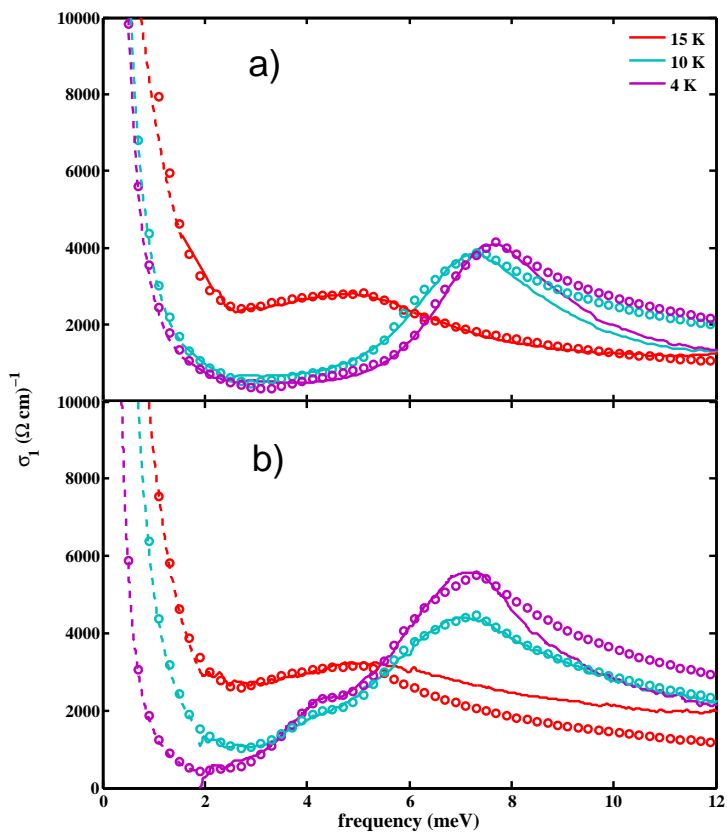}}
\vspace*{0.0 cm}%
\caption{(color online): Fits of a Dynes model for the density of states to the optical conductivity for the a-axis (top panel) and c-axis (bottom panel) for selected temperatures in the hidden order state. The dashed lines at low frequency indicate the fit to a Drude model. Circles are experimental points, the lines are the fits. After, J.S. Hall {\it et al.}, Phys. Rev. B {\bf 86} (2012) p.035132.}
\label{fig:sigma.ho}
\vspace*{0.0cm}
\end{center}
\end{figure*}

Despite this, the optical signature of the hidden order state observable in the conductivity shown in Fig. 7 from Hall {\it et al.}\cite{hall12}is strongly reminiscent of a spin density wave state for light polarized parallel to both the a-axis and the c-axis. It was initially noted that the characteristic structure of a suppression of conductivity below the gap energy with spectral weight recovery in a symmetric peak at higher frequency bore a striking resemblance to the spin density wave state in Cr \cite{bonn88, barker68}. At the time, theoretical descriptions of the SDW conductivity for materials in which the scattering rate is larger than the gap energy were not available, but an estimate could be made for the optical energy gap $2\Delta$ of between 5.5 meV and 8 meV \cite{bonn88}. This was smaller than the specific heat value of 11 meV \cite{maple86}, a first indication that optical measurements and specific heat would come to disagree. In weak-coupling BCS theory there is a universal relationship between the energy of the gap and the transition temperature, given by $2\Delta=3.53k_{B}T_{c}$. A transition temperature of 17.5 K implies a gap energy of $2\Delta = 5.3$ meV.

Recent measurements of the optical conductivity of URu$_2$Si$_2$ \cite{hall12} reveal the anisotropy of the hidden order parameter in both the different magnitudes of the optical gaps when measured with light polarized along either the a- or the c-axis, but also in the appearance of a second gap, presumably on a different part of the Fermi surface. The strong suppression of the conductivity indicates that a substantial portion of the Fermi surface is being gapped. We know from quantum oscillations \cite{hassinger11} that there are four or five Fermi surface sheets in the hidden order state, so presumably the gaps affect each sheet differently. Optical spectroscopy is not a momentum resolved probe, as it is limited to constant ${\bf k}$-vector transitions and averaging over all of k-space for ${\bf k||E}$ and the individual components of the Fermi surface cannot separated.

It is possible to fit the measured optical conductivity \cite{hall12} with a simple Dynes \cite{dynes78} model for an s-wave gap in the density of states:

\begin{eqnarray} 
n_{D}(E)=|Re\frac{E/\Delta + i\gamma}{\sqrt{(E/\Delta + i\gamma)^2-1}}| 
\end{eqnarray} 

\noindent where a factor of $\gamma$ has been introduced to account for a finite quasiparticle lifetime, but it can also account for some anisotropy in the gap. When impurity scattering, the frequency-dependent effective mass, and finite quasiparticle lifetime effects are included, the singularity $ n_{D}$ is broadened and the density-wave peak in $\sigma_{1}(\omega)$ takes on the characteristic shape seen in Cr \cite{barker68, gruner94} and the Bechgaard salt (TMTSF)$_2$PF$_6$ \cite{degiorgi96}. The optical conductivity is given by integrating the joint density of states:

\begin{eqnarray}
\sigma_D(\omega) = Re\frac{1}{\omega}\int_{\Delta}^{\omega_c}n_D(\omega')n_D(\omega-\omega') d\omega'
\end{eqnarray}

\noindent In a broken symmetry ground state such as superconductivity or a spin density wave, there are two possible transition processes between any two quasiparticle states. The transition probabilities are determined by coherence factors, which depend on whether the two transition processes interfere with one another destructively (case I) or constructively (case II) \cite{tinkham96, dressel02}. The BCS formalism describes both density wave states (case I coherence factor) and superconductivity (case II coherence factor) depending on whether the effective interaction between the quasiparticles changes sign on opposite sides of the Fermi surface (that is, when going from ${\bf k}$ to -${\bf k}$). The model above uses a simplified Dynes density of states with broadening to reproduce the qualitative features of the conductivity and extract a reasonable estimate of the gap in the hidden order state. It has been noted \cite{guo12} that URu$_2$Si$_2$ is an almost archetypical example of a case I coherence factor. Indeed, a theoretical calculation of the optical response of heavy Fermion spin density wave materials \cite{dolgov02} noted that, of all of the candidates, URu$_2$Si$_2$ was the only material that perfectly matched the calculated optical conductivity.

In many ways, this makes a great deal of sense. The gapping of incommensurate magnetic excitations at the hidden order transition has been shown by neutron scattering \cite{broholm87, wiebe07}, and these account for much of the entropy lost. Band structure calculations \cite{elgazzar09, oppeneer11} have yielded a picture of the Fermi surface with strong nesting in the pressure-induced anti-ferromagnetic state, and quantum oscillation measurements \cite{hassinger11} demonstrate that there is no significant Fermi surface restructuring between HO and LM-AFM, which implies that the Fermi surface calculated for the latter state applies equally well to the former. Incommensurate nesting will lead to the formation of a spin-density wave gap at $\epsilon_{F}$ and will be accompanied by a sharp absorption feature in the optical data \cite{lee73, degiorgi97}, while a commensurate antiferromagnetic order of the localized moments would not be visible in optical measurements because it would not lead to a gap in the excitation spectrum at $\epsilon_{F}$.

An analysis of the spectral weight $N_{eff}$ transfer in the hidden order region \cite{guo12} tells the story of what happens as the material transitions from the incoherent to the coherent and then hidden order state. The hybridization transfers spectral weight to higher frequencies, above 300 meV. In contrast, the hidden order transfers spectral weight into the peak immediately above the gap region. However, the spectral weight above the peak in the HO state is not equal to the spectral weight in the coherent state at 20 K. This effect is difficult to quantify precisely, however, as much depends on the choice of plasma frequency for the Drude weight and how the Drude peak is determined below the measurement region (typically $\sim$ 2.5 meV).

\begin{figure*}
\begin{center}
\vspace*{-0.5 cm}
\hspace*{-0.8cm}
\resizebox*{10cm}{!}{\includegraphics{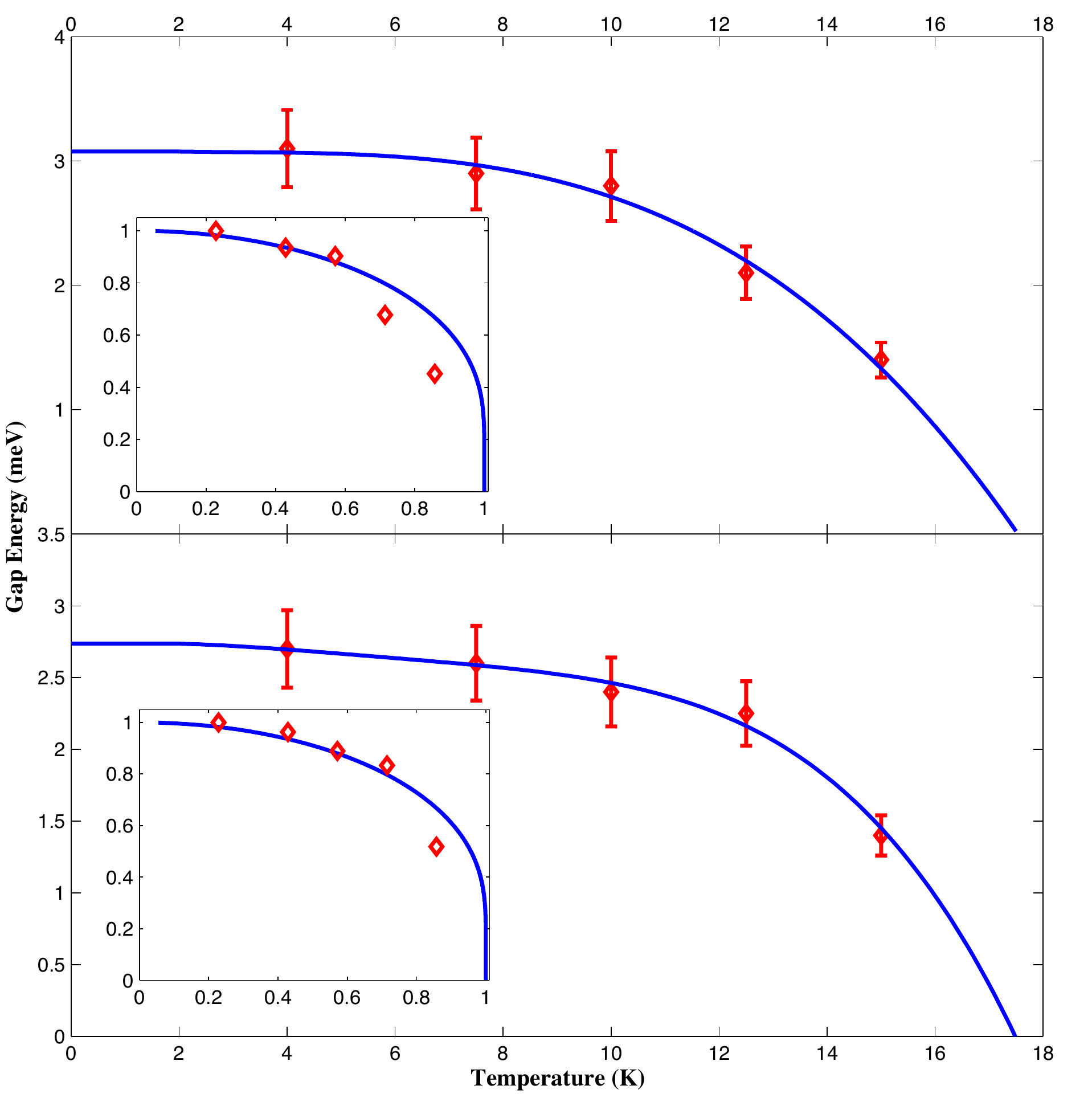}}
\vspace*{0.0 cm}%
\caption{(color online): The temperature dependence of the gap $\Delta$ in the hidden order state as a function of temperature.  The upper panel shows {\bf E} parallel to the ab plane; the lower panel, {\bf E} parallel to the c-axis. The inset is the mean field temperature dependence $2\Delta = 3.5 k_BT_{HO}$, the solid line is a guide to the eye. After, J.S. Hall {\it et al.}, Phys. Rev. B {\bf 86} (2012) p.035132.}
\label{fig:delta.ho}
\vspace*{0.0cm}
\end{center}
\end{figure*}

The temperature dependence of the gap parameter $\Delta$ as determined by fits to the Dynes function is shown in figure \ref{fig:delta.ho}.
The upper panel shows the $ab$-plane gap and the lower panel  the $c$-axis (larger) gap. The solid lines are guides to the eye, extrapolated to zero at the temperature of the HO transition at 17.5 K.  The inset shows the expected mean field theory gap energy $2\Delta = 3.5 k_BT_{HO}$.  It is clear that the data do not support the mean field dependence.  It must be noted, however, that the presence of multiple gaps of different magnitudes can lead to deviations from the mean field temperature dependence of the order parameter\cite{nicol05}. Such a scenario may account for the temperature dependence of the hidden order gap in \ur.

The limiting value at low temperature from Hall \etal\cite{hall12} for the $ab$-plane gap of $\Delta_{ab}=3.2$ meV is in reasonable agreement with the work of other investigators if one takes into account possible variations that can be attributed to different criteria for the gap location. Two gaps are seen for $E || c$ at 2.7 meV and a lower one at 1.8 meV. Bonn gives a range from  $\Delta$ = 2.8 meV to 4.0 meV \cite{bonn88}, while Guo \etal give $\Delta$ = 4 meV \cite{guo12} . Other techniques yield gaps that have a larger range of variation, in part due to the fact that their resolution is lower than what is common in optics.  Both  STM \cite{aynajian10} and ARPES \cite{santandersyro09, boariu13} also give gap values of about 4 meV as well as neutron scattering (around 4 meV) \cite{wiebe07,williams12}.  Other tunneling data include those of Escadero \etal\cite{escadero94} who find $\Delta= 5.85$ meV. Another way to find the gap is to fit a Boltzmann factor to various thermodynamic measurements.  We will assume that the gap fitted there is $2\Delta$.  Thus the earliest work of Palstra \etal \cite{palstra85}, who fit the specific heat coefficient to an activation energy, reports a gap of 115 K {\it i.e.} in our notation, $\Delta=5$ meV. Mentink \etal\cite{mentink96} fit the dc resistivity in the hidden order state to an activation model and get an $ab$-plane gap of 3.1 meV and $c$-axis gap of 2.2 meV.

In summary, the optical data point to a hidden order state with multiple anisotropic gaps.  The non-mean field temperature dependance is in accord with this. The magnitude of the gap parameter $\Delta$ agrees with other spectroscopic measurements if one takes into account the lower resolution of ARPES and STM techniques.  Overall, the spectrum can be fit closely by a Dynes model with finite quasiparticle lifetimes, and is modeled accurately\cite{dolgov02} by a BCS model using type I coherence factors for a spin density wave gap. The transfer of spectral weight is very similar to what one sees in density wave transitions, although one has to rule out a simple SDW picture for the lack of a sufficient ordered moment.

\vspace{-0.5cm}

\section{Summary and Conclusion}

We will now try to bring together all the optical data keeping in mind the other spectroscopic results as well.  It is natural to discuss the results in terms of the temperature regions defined in figure \ref{fig:dc.res} but we have to modify the original picture by including a separate "precursor" region between 30 K and $T_{HO}$.

Between 300 K and 30 K \ur behaves very much like other heavy fermion materials.  Above $T_K$ the conductivity is incoherent and increases slightly as the temperature is lowered, reaching a broad maximum around 70 K, and then drops smoothly.  We have identified this decreasing conductivity as a combination of $f$-electron hybridization and a general reduction of scattering due to thermal factors.  The overall result is a growing Drude peak and a hybridization gap at 10 meV.  However, there remains a substantial incoherent background even at 17.5 K and the hybridization gap is not complete: the conductivity at the minimum at 10 meV has only dropped to half its room temperature value.  The scattering mechanism of the Drude component is Fermi-liquid-like in that the frequency dependence is $\omega^2$, but the scattering is anomalous in that is does not follow the scaling expected for electron-electron umklapp scattering.  A possible mechanism is resonant scattering from the remaining un-hybridized $f$-electrons. There are no dominant features in the optical spectra from bosonic interactions as seen in the cuprate\cite{carbotte11} and the pnictide\cite{yang09} superconductors in the normal state above their superconducting transition temperatures. 

Below 30 K, several things happen. There is a break in the rate of change of the Drude plasma frequency, and the rate of change of the dc resistivity begins to deviate from linearity. As the temperature is lowered further, time resolved THz spectroscopy, point contract spectroscopy, and ARPES all show evidence for the opening of a partial gap in the electronic excitation spectrum.  None of the measured effects that begin between 30 K and $T_{HO}$ are discontinuous at the hidden order transition and ARPES sees a gap open on a different part of the Fermi surface at $T_{HO}$. This strongly implies that the changes in the electronic structure in  this temperature range are unrelated to the hidden order state.  

Below $T_{HO}$ several things happen.  A gap opens up and within a fraction of a degree Kelvin the resistivity increases due to a loss of states.  But as the temperature is lowered the reduced scattering becomes more important and the resistivity drops exponentially. A gap in the optical conductivity opens up and the gapped spectral weight is transferred to a narrow band just above the gap. It should be noted that it is the {\it incoherent} non-Drude part of the conductivity that is gapped, but above 12 meV the conductivity remains at its normal state value.  This is consistent with hybridization stopping at the hidden order transition while above the frequency of the hidden order gap the carriers remained incoherent. While these dramatic changes take place in the incoherent channel of the conductivity, the Drude peak narrows but does not show any large discontinuous changes in width at the hidden order transition.     
\vspace{-0.5 cm}
\section{Acknowledgements}

We thank the following colleagues for helpful discussions: Bill Buyers, Jules Carbotte, Andre Chubukov, Piers Coleman, Seamus Davis, Gabriel Kotliar, Brian Maple, Dimitrii Maslov and Peter Oppeneer.  In particular we owe thanks to Nick Butch for allowing us to use his high resolution resistivity data shown in figures \ref{fig:dc.res} and \ref{fig:drho}. Much of the work shown here was done in collaboration with Urmas Nagel and Toomas R{\~o}{\~o}m in Tallinn, Estonia and Ricardo Lobo in Paris. This work was supported by the Natural Sciences and Engineering Research Council of Canada and the Canadian Institute for Advanced Research.

\bibliographystyle{tPHM}
\bibliography{PhilMag4}

\end{document}